\documentclass[12pt,preprint]{aastex}
\def\jcap{JCAP}
\usepackage{xcolor}

\begin{document}
\title{An Observed Evidence for the Primordial Origin of Galaxy Sizes}
\author{Jun-Sung Moon$^{1,2,3}$ and Jounghun Lee$^{1}$}
\affil{$^1$Department of Physics and Astronomy, Seoul National University, Seoul 08826, Republic of Korea}
\affil{$^2$Research Institute of Basic Sciences, Seoul National University, Seoul 08826, Republic of Korea}
\affil{$^3$Institute of Astronomy and Astrophysics, Academia Sinica, 11F of Astronomy-Mathematics Building, No. 1, Sec. 4, Roosevelt Rd., Taipei 106319, Taiwan, R.O.C.}
\email{jsmoon.astro@gmail.com, cosmos.hun@gmail.com}
\begin{abstract}
We present an observational evidence supporting the scenario that the protogalactic angular momenta play an important role in 
molding the optical sizes of present galaxies. Analyzing the NASA-Sloan Atlas catalog in the redshift range of $0.02\le z<0.09$, 
we observationally determine the probability density distributions, $p(r_{50})$ and $p(r_{90})$, where $r_{50}$ and 
$r_{90}$ denote the galaxy sizes enclosing $50\%$ and $90\%$ of their $r$-band luminosities, respectively. 
Both of the distributions are found to be well described by a bimodal Gamma mixture model, which is consistent with the recent numerical results.  
Classifying the local galaxies by their ratios, $r_{50}/r_{90}$,  we also show that for the case of late-type galaxies with $r_{50}/r_{90}\ge 0.45$ 
both of $p(r_{50})$ and $p(r_{90})$ exhibit no bimodal feature, following a unimodal Gamma model. 
Assuming the existence of a linear causal correlation between $\{r_{50},r_{90}\}$ of the late-type galaxies and the primordial spin factor, $\tau$, 
defined as the degree of misalignments between the initial tidal and protogalaxy inertia tensors, 
we reconstruct the probability density distributions, $p(\tau)$, directly from the observationally determined $p(r_{50})$ and $p(r_{90})$ 
of the late-type galaxies. It is shown that the reconstructed $p(\tau)$ is in an excellent agreement with the real distribution of $\tau$ that was 
determined at the protogalactic stages by numerical experiments. A critical implication of our result on reconstructing the initial conditions 
from observable galaxy sizes is discussed.
\end{abstract}
\keywords{Unified Astronomy Thesaurus concepts: Cosmology (343); Large-scale structure of the universe (902)}
\section{Introduction}\label{sec:intro}

The galaxy angular momentum is one of the most fundamental properties that the galaxies possess, the origin of which is believed to be 
the protogalactic tidal interactions with the surrounding matter distribution \citep{dor70,whi84}. 
Until the primordial protogalactic sites decouple from the surrounding matter distribution at the turn-around moments,  they steadily acquire angular 
momenta under the tidal influences, developing causal correlations with the initial tidal fields on the scale of their total masses~\citep{CT96,LP00,LP01}. 
If the angular momenta were perfectly conserved after turning around, then both of the magnitudes and directions of the angular momenta of the present galaxies 
would exhibit strong correlations with the initial conditions. The directions of the present galaxy angular momenta would exhibit strong alignments with the principal 
axes of the linear tidal fields, while their magnitudes would show strong correlations with those of protogalactic angular momenta, which were found to be 
proportional to the degree of alignments between the principal axes of linear density and potential hessian tensors~\citep{ML25}. 

It used to be thought that  the galaxy angular momenta would be prone to severe modifications during the subsequent nonlinear evolutions, losing much of 
their memory of the protogalactic states \citep[e.g.,][]{por-etal02a,por-etal02b,vit-etal02}.  Several numerical experiments based on hydrodynamical simulations indeed 
reported their failures in finding any significant alignments between the directions of stellar and dark matter (DM) angular momenta of galactic 
halos~\citep[e.g.,][and references therein]{jia-etal19}.  Absence of correlations between the DM and stellar angular momenta of galactic halos 
were ascribed to the vulnerability of baryonic components to complicated non-gravitational astrophysical processes associated with gas inflow and star formation, which 
implied no existence of any connections between observable stellar angular momenta of present galaxies and initial tidal fields.

However, recent numerical and observational studies have revealed that the stellar angular momenta of present galaxies still retain non-negligible 
correlations not only with their DM angular momenta but also with the initial conditions~\citep{tek-etal15,she-etal23,she-etal24}. 
For instance, \citet{mot-etal21} detected a $\sim 2.7\sigma$ signal of correlations between spin directions of local 
spiral galaxies extracted from the Galaxy Zoo data~\citep{zoo1,zoo2} and principal axes of the linear tidal fields reconstructed by the ELUCID 
collaborations~\citep{elucid1,elucid2}. 
\citet{cad-etal22} performed multiple zoom-in hydrodynamical simulations and showed that the protogalactic angular momenta have causal effects  
on the magnitudes of stellar angular momenta of present galactic halos by controlling their merging processes. 
\citet{ML23} found by analyzing the galactic halos identified in a high-resolution cosmological hydrodynamical simulation that their stellar angular momenta 
exhibit significantly strong alignments with the principal axes of the initial tidal fields and that the alignment strengths depend on the degree of misalignments 
between initial tidal and protogalactic inertia tensors. 
These numerical and observational evidences have supported the theoretical claim that the galaxy angular momenta can in principle be used to reconstruct the 
initial tidal and density fields~\citep{LP00,LP01} and to constrain the early universe physics like large-scale parity violation, neutrino masses, and dark energy 
equation of state \citep{yu-etal19,yu-etal20,lee-etal20,LL20,shi-etal24}. 

Two different approaches have been taken in the previous studies to investigate the connections between galaxy angular momenta and initial 
conditions. In one approach the directions of galaxy angular momenta and their alignments with the principal axes of initial tidal fields are the primary subjects. 
Meanwhile the other approach concerns the galaxy spin parameters~\citep{bul-etal01} and their dependence on the magnitudes of protogalaxy angular momenta.  
The two approaches have their own merits and drawbacks, complementing each other and providing independent information on the effects of initial conditions 
on galaxy angular momenta. 
The best merit of the first approach is its observational feasibility.  Under the key assumption that the observable angular momenta of galaxy stellar components 
are well aligned with those of dark matter counterparts, the spin axes of spiral galaxies can be determined with two-fold degeneracy 
from information on their position and inclination angles as well as their axial ratios~\citep[e.g.][]{lee11}. 
Recent numerical works, however, casted a doubt on this key assumption by showing that the stellar angular momenta of spiral galaxies exhibit not only 
misalignments with those of DM counterparts but also quite a different tendency of alignments with the principal axes of initial tidal fields~\citep{ML23}. 
Besides, this approach requires a valid reconstruction of the initial tidal field on the proto-galactic scale of $\sim 1$-$2\,h^{-1}{\rm Mpc}$,  
which is difficult to achieve in practice~\citep{mot-etal21}. 

The second approach based on the magnitudes of galaxy angular momenta has been regarded as being less attainable due to the practical 
difficulty in determining the galaxy spin parameters from real observations.  A clue about how to overcome this difficulty, however, has recently been found in the work of  
\citet{ML25}, according to which the observable galaxy stellar sizes share significant amount of mutual information with the primordial spin factors that are directly 
proportional to the magnitudes of protogalactic angular momenta. Their results implied that instead of the galaxy spin parameters 
the observable galaxy sizes can be used to investigate how much memory the angular momenta of present galaxies have for the magnitudes of 
protogalaxy angular momenta. 

In this Letter, we attempt to find an observational evidence supporting that the primordial spin factors may have a causal effect on the galaxy optical sizes  
by reconstructing the probability density distributions of the former from those of the observable latter. 
In Section~\ref{sec:psize} is presented a detailed observational analysis via which we determine the probability density distributions of the observed galaxy sizes, 
which will be used to reconstruct the probability density distributions of the primordial spin factor. 
In Section~\ref{sec:ptau}, we explain how the probability density distributions of the observed galaxy sizes are statistically converted to those of the primordial spin 
factors and present a comparison of the converted distributions with the numerical results obtained in the prior work.
In Section~\ref{sec:con}, we summarize the results and discuss its physical implication.

\section{Observational Determination of the Galaxy Size Distributions}\label{sec:psize}

Our analysis utilizes the NASA-Sloan Atlas catalog that compiles the photometric images and shape parameters of local galaxies~\citep{bla-etal11} 
observed by the Sloan Digital Sky Survey~\citep{sdssdr12,sdssdr13} and Galaxy Evolution Explorer~\citep{galax}. 
The catalog includes information on the galaxy half-light and $90\%$-light sizes ($r_{50}$ and $r_{90}$, respectively) determined in the $r$-band 
via the most reliable {\it elliptical Petrosian aperture photometry}~\citep{wak-etal17}. 
For a proper determination of the probability distributions, $p(r_{50})$ and $p(r_{90})$, of the NASA-Sloan Atlas galaxies,  it is necessary 
to exclude those galaxies with angular sizes smaller than the photometric seeing (FWHM) since their values of $r_{50}$ and $r_{90}$ are likely to be 
overestimated ~\citep{she-etal03,mas-etal10}.  

\begin{figure}[ht]
\centering
\includegraphics[height=6cm,width=16cm]{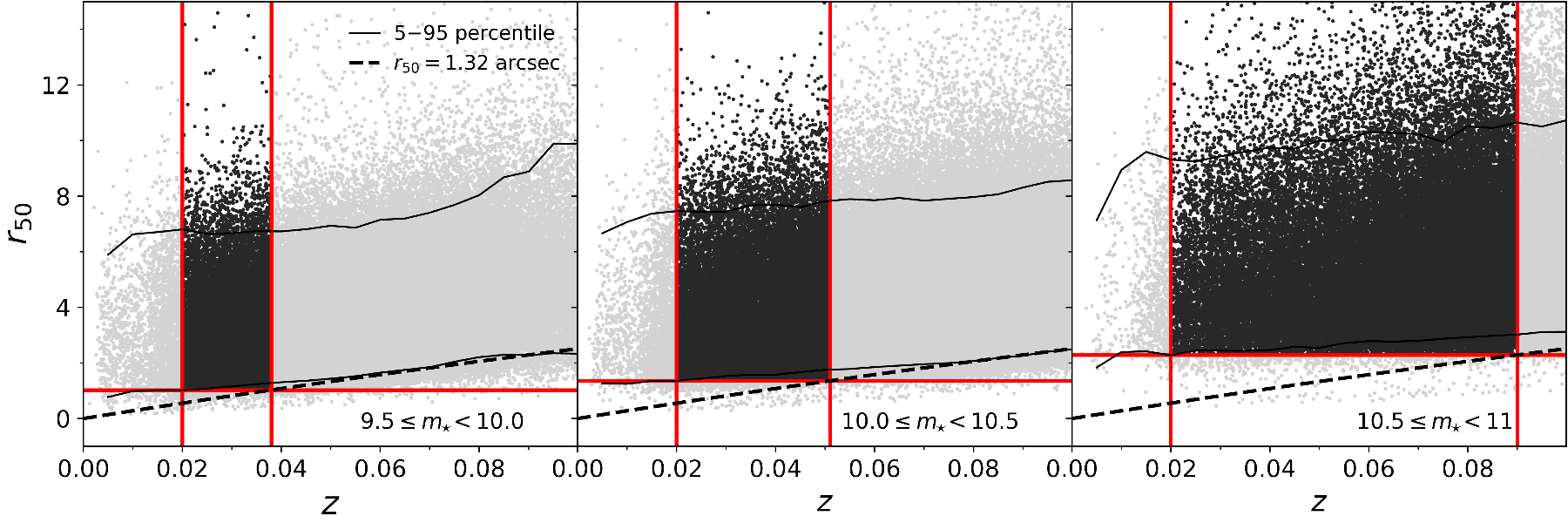}
\caption{Scatter plots of the redshifts and the half-light sizes of the NASA-Sloan-Atlas galaxies in three different stellar-mass ranges, 
with the locations of two redshift cutoffs, $z_{\rm min}$ and $z_{\rm max}$ (vertical red lines) applied to exclude those galaxies 
the angular sizes of which are smaller than the photometry seeing at $z_{\rm max}$ (horizontal red lines).}
\label{fig:zmax}
\end{figure}

Given that the angular sizes of local galaxies vary not only with redshifts but also with stellar masses, $M_{\star}$, we split the ranges 
of $m_{\star}\equiv \log\left[M_{\star}/(h^{-1}M_{\odot})\right]$ into three intervals:  $[9.5,\ 10)$, $[10,\ 10.5)$ and $[10.5,\ 11]$ and find the galaxies 
belonging into each $m_{\star}$-bin at redshifts $z\ge z_{\rm min} = 0.02$, using information provided by the NASA-Sloan Atlas catalog information on $M_{\star}$.   
We set a minimum redshift $z_{\rm min}$ because large galaxies at low redshifts can be excluded from the spectroscopic sample due to their very bright apparent 
magnitudes.
Then, in each $m_{\star}$-interval,  we take the following three steps to determine the maximum redshift, $z_{\rm max}$, beyond which the galaxy 
angular sizes drop below the photometric seeing.
\begin{itemize}
\item
Among the galaxies belonging to the upper $95\%$ in the values of $r_{50}$ at $z_{\rm min}$, find its lowest value, $r_{50,{\rm min}}$, which serves as the lower limit of $r_{50}$ for galaxies within the mass bin.
\item
Determine the maximum redshift, $z_{\rm max}$, at which $r_{50,{\rm min}}$ matches the SDSS median photometric seeing (1.32 arcsec in $r$-band). 
\item
Select only those galaxies which satisfy two conditions of $r_{50} \ge r_{50,{\rm min}}$ and $z_{\rm min} \le z \le z_{\rm max}$. 
\end{itemize}
Figure~\ref{fig:zmax} shows the redshift cutoff values (red vertical lines) determined via the above procedure in each $m_{\star}$-interval. 
Through this procedure, we create three size-limited samples containing only those NASA-Sloan Atlas galaxies located at redshifts lower than 
$z_{\rm max}$ to obtain reliable size distributions.  A total of $13370$, $26783$ and $58514$ galaxies are found to be included in the three 
controlled samples corresponding to the three $m_{\star}$-ranges in an increasing order. 

\begin{figure}[ht]
\centering
\includegraphics[height=15cm,width=16cm]{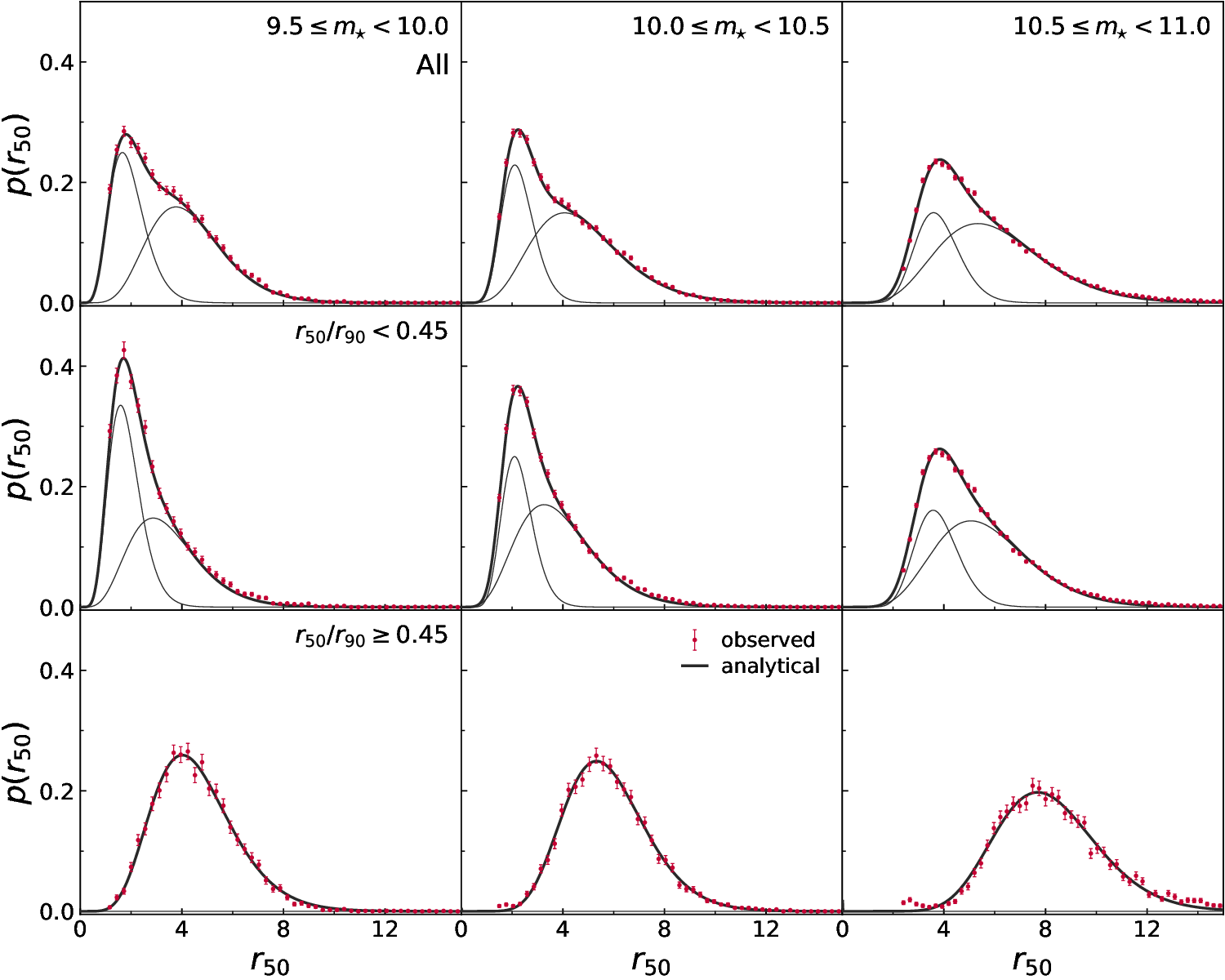}
\caption{Probability density distributions of the half-light sizes of the selected local galaxies (red filled circles with Poisson errors) 
from the NASA-Sloan-Atlas catalog along with the best-fit Gamma mixture model (black thick solid lines) composed of two distinct modes 
(black thin solid lines) in three different stellar-mass ranges for three different cases of galaxy light concentrations, $r_{50}/r_{90}$.}
\label{fig:pro50}
\end{figure}

For each sample, we split the range of $r_{50}$ into multiple bins of equal length, $\Delta r_{50}$, and count the number, $\Delta N$, of those galaxies whose 
half-light sizes fall in each $r_{50}$-bin. The probability density, $p(r_{50})$, at each $r_{50}$-bin is computed as $p(r_{50}) = \Delta N/(N\Delta r_{50})$. 
Figure~\ref{fig:pro50} plots $p(r_{50})$ (filled red circles) with Poisson errors from each of the three samples in the top panels, 
clearly demonstrating its bimodal feature in all of the three $m_{\star}$-ranges. This observational result is consistent with the recent 
numerical finding that the probability density distributions of stellar sizes are very well described by the following Gamma mixture model~\citep{ML25}: 
\begin{equation}
\label{eqn:bim}
p(r) = \xi\,p\left(r; k_{1},\theta_{1}\right) + (1-\xi)p\left(r; k_{2},\theta_{2}\right)\, . 
\end{equation}
Here, $p(r;k,\theta)$ is the Gamma distribution defined as
\begin{equation}
\label{eqn:gam}
p\left(r; k,\theta\right) =\frac{1}{\Gamma(k)\theta^{k}}r^{k-1}\exp\left(-\frac{r}{\theta}\right)\, ,
\end{equation}
with Gamma function $\Gamma(k)$ and two adjustable parameters $\{k,\theta\}$. Note that the bimodal Gamma mixture model given in 
Equation~(\ref{eqn:bim}) is characterized by five adjustable parameters, $\{k_{i},\theta_{i}\}_{i=1}^{2}$ and $\xi$, where $\xi$ denotes the fraction of two 
Gamma modes. 

Fitting Equation~(\ref{eqn:bim}) to the observationally obtained $p(r_{50})$, we determine the best-fit values of $\{k_{i},\theta_{i}\}_{i=1}^{2}$ and $\xi$ 
with the help of the $\chi^{2}$-statistics. 
Figure~\ref{fig:pro50} shows the best-fit Gamma mixture model for $p(r_{50})$ (black thick lines), revealing good agreements between the 
analytic models and numerical results in all of the three $m_{\star}$-ranges. Note that the bimodal feature of $p(r_{50})$ becomes less conspicuous in a 
higher $m_{\star}$-range. 
Given that the galaxies with higher $m_{\star}$ tend to consist of early-type galaxies, this result implies a possible correlation between galaxy morphology and the 
bimodality of galaxy size distributions.

To explore this implication further, we classify the galaxies in each $m_{\star}$-range into two subsamples according to the ratios, $r_{50}/r_{90}$, 
a measure of their concentrations that are known to reflect quite well their morphologies~\citep[][and references therein]{oh-etal13}.  
The threshold of this ratio is set at the conservative value of $0.45$ \citep{PC05}, which is often adopted to segregate late-type galaxies ($r_{50}/r_{90}\ge0.45$) 
from early-type ones ($r_{50}/r_{90} < 0.45$). Then, we separately determine $p(r_{50})$ from each subsample in each $m_{\star}$-range and find the 
best-fit parameters of the Gaussian mixture model with the help of the $\chi^{2}$-statistics, the results of which are shown in the middle and bottom panels of 
Figure~\ref{fig:pro50}. 

\begin{figure}[ht]
\centering
\includegraphics[height=15cm,width=16cm]{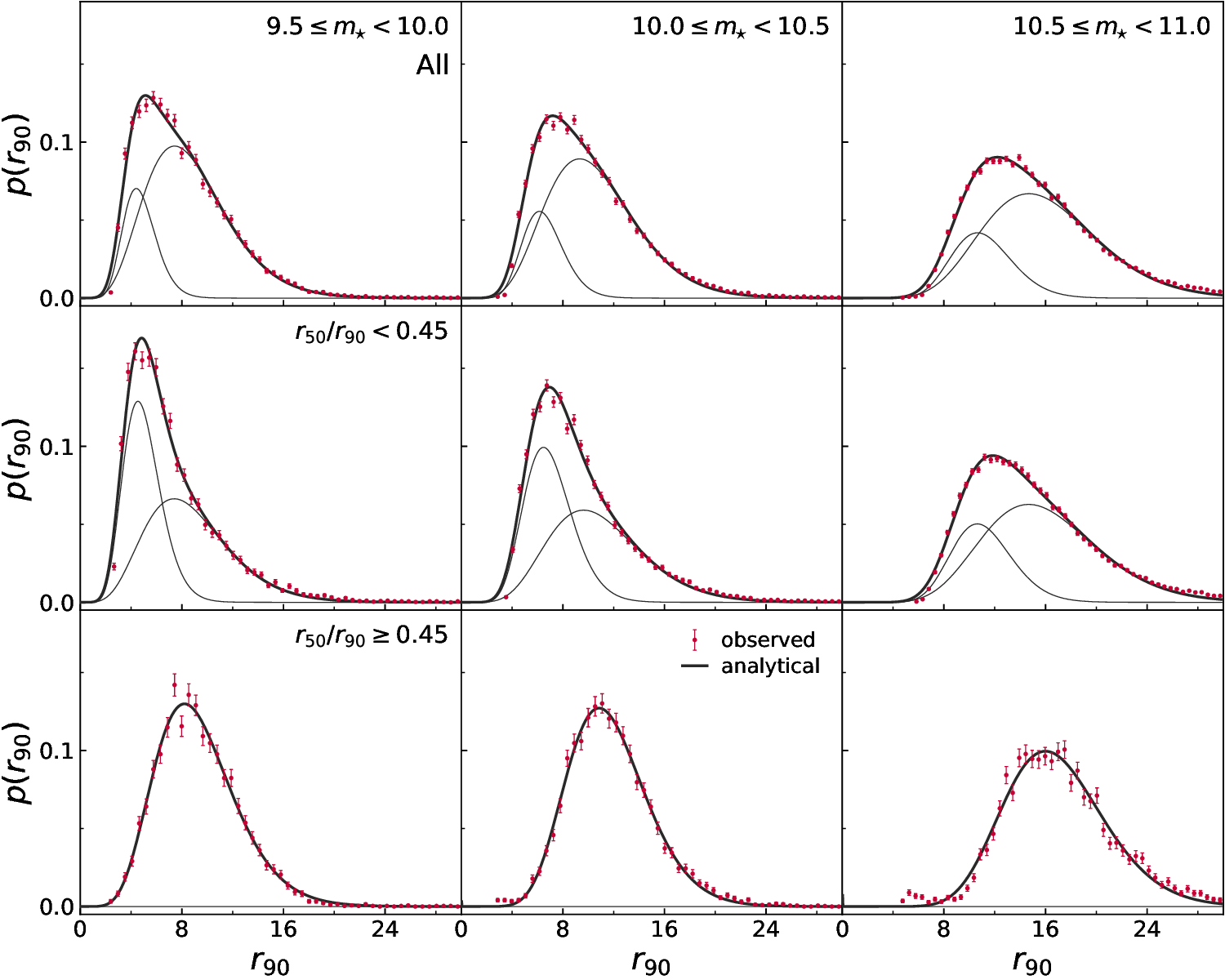}
\caption{Same as Fig.~\ref{fig:pro50} but of the $90\%$-light sizes.}
\label{fig:pro90}
\end{figure}

As can be seen, the early-type galaxies with $r_{50}/r_{90} < 0.45$ exhibit highly bimodal distributions (middle panels), yielding the best-fit value of $\xi\sim 0.5$ in each 
$m_{\star}$-range.  Whereas, for the case of late-type galaxies (bottom panels), the size distributions show no bimodal feature ($\xi=0$), being well described 
by the {\it unimodal} Gamma distribution (Eq.~[\ref{eqn:gam}]) alone in all of the three $m_{\star}$-ranges. 
We repeat the whole process but with the $90\%$-light sizes and show the results in Figure~\ref{fig:pro90}.  As can be seen the behaviors and trends of 
$p(r_{90})$ are very similar to those of $p(r_{50})$. Especially, the degree of bimodality of $p(r_{90})$ is also strongly dependent on the ratio, 
$r_{50}/r_{90}$, becoming negligibly low for the late-type case just like that of $p(r_{50})$.  These results imply that the galaxy optical size distributions 
must develop their bimodal features during the relaxation processes after their latest major merger events. 

Given the previous numerical findings that the spin parameters of galactic halos are found to have the strongest causal 
correlations with protogalactic angular momenta~\citep{ML24a} and to follow the unimodal Gamma distribution~\citep{ML24b}, 
another critical implication of the results shown in the bottom panels of Figures~\ref{fig:pro50}--\ref{fig:pro90} is that 
for the investigation of correlations between galaxy optical sizes and protogalactic angular momenta, the best target should be the 
late-type galaxies with $r_{50}/r_{90}> 0.45$.  If the protogalactic angular momenta are indeed causally correlated with the galaxy optical sizes, 
then the probability density distributions of two quantities could be converted to each other. In other words, if the probability density distributions 
converted from the observed galaxy optical size distributions turn out to reproduce the behaviors and shapes of real protogalactic angular momenta 
distributions determined in the numerical experiments, it will observationally support the existence of such causal correlations. 

Hereafter, we will focus only on the late-type galaxies and use their optical size distributions to reconstruct the distributions of 
protogalactic angular momenta.  Table~\ref{tab:fit} provides information on the number of late-type galaxies (second column) 
the best-fit parameters of the unimodal Gamma model (third and fourth columns), and the minimum physical sizes (seventh column) 
in the three $m_{\star}$-ranges. The values inside the parentheses correspond to the $90\%$-light sizes. 
 
\section{Reconstruction of the Primordial Spin Factor Distributions}\label{sec:ptau}

According to the linear tidal torque theory~\citep{dor70,whi84,CT96,LP00}, the protogalactic angular momenta are directly proportional to the 
degree of misalignments between the principal axes of two tensors, $(I_{ij})$ and $(T_{ij})$ (protogalactic inertia and initial tidal tensors, respectively), 
which \citet{ML24a} quantified as
\begin{equation}
\label{eqn:tau}
\tau \equiv \left(\frac{I^{2}_{12}+I^{2}_{23}+I^{2}_{31}}{I^{2}_{11}+I^{2}_{22}+I^{2}_{33}}\right)^{1/2}\, ,
\end{equation}
in the principal frame of $(T_{ij})$.  \citet{ML24b} numerically derived its probability density distribution, $p(\tau)$,  and found it to be 
well approximated by the unimodal Gamma model (Eq.~[\ref{eqn:gam}]), regardless of the scales used to smooth $(T_{ij})$. 
Noting that $p(\tau)$ tends to have broader shapes on larger scales,  they claimed that this scale dependence of $p(\tau)$ represent 
multi-scale tidal influences on the protogalactic angular momenta. 

In the follow-up work of \citet{ML25} based on high-resolution hydrodynamical simulations, it has been revealed that the primordial spin factor, $\tau$, shares 
significantly large amount of mutual information with the stellar-mass sizes of galactic halos and that their correlations are almost linear, albeit with 
large scatters~\citep[see Figure 5 in][]{ML25}. Given the result of \citet{ML25}, we model $\tau$ as 
\begin{equation}
\label{eqn:tau_re}
\tau = \alpha\left(r - r_{\rm min}\right) ,\qquad {\rm for}\,\,\, r\in\{r_{50},r_{90}\}\, ,
\end{equation}
where $\alpha$ is the slope of the linear correlation between $r$ and $\tau$.  
To statistically take into account the scatters in the linear correlations between $r$ and $\tau$, we treat $\alpha$ as a Gaussian 
random variable with mean $\langle\alpha\rangle$ and standard deviation $\sigma_{\alpha}$. 

\begin{figure}[ht]
\centering
\includegraphics[height=15cm,width=15cm]{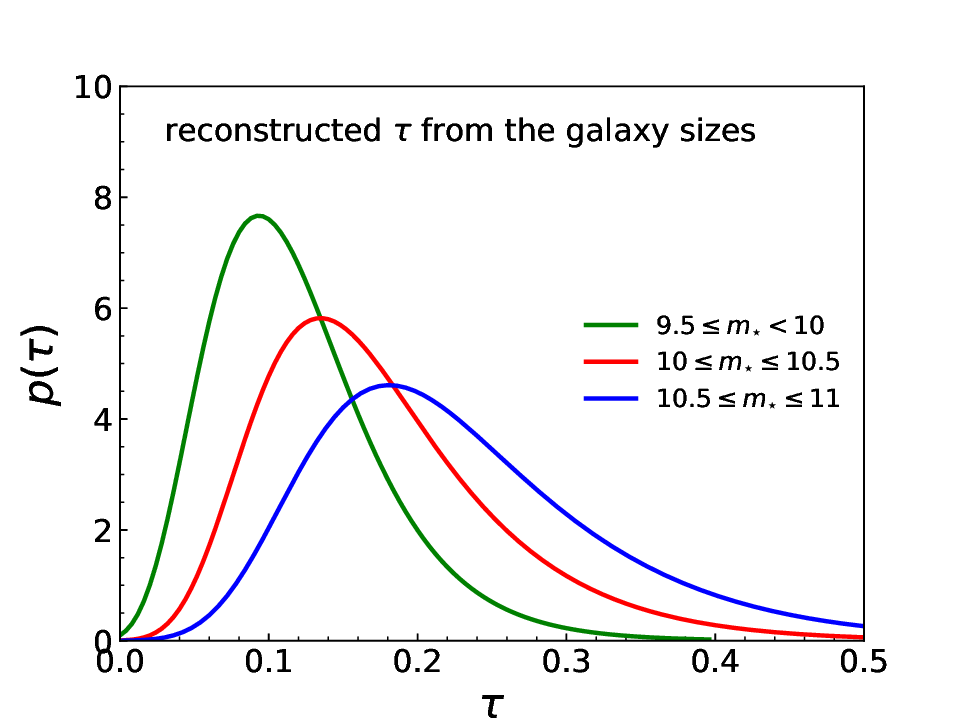}
\caption{Reconstructed probability density distributions of the primordial spin factor from those of the observed galaxy optical sizes 
via Eq.~(\ref{eqn:ptau_re}) in three different $m_{\star}$ ranges. The reconstructed distributions match the real ones determined 
by numerical simulations directly from protogalactic inertia and initial tidal fields smoothed on three different scales.}
\label{fig:ptau_re}
\end{figure}

Equation~(\ref{eqn:tau_re}) enables us to reconstruct the probability density distribution of $\tau$ from the Gamma distributions that the observed 
size distributions of the late-type galaxies are found to follow in Section~\ref{sec:psize}:
\begin{eqnarray}
p(\tau) &=& \int_{0}^{\infty} p\left(r\left[\tau\right],\alpha\right)d\alpha =  \int_{0}^{\infty} p\left(\frac{\tau}{\alpha}+r_{\rm min},\alpha\right)d\alpha \nonumber \\
&=& 
\label{eqn:ptau_re}
\frac{1}{\sqrt{2\pi}\sigma_{\alpha}}\frac{1}{\Gamma(k)\theta^{k}}\int_{0}^{\infty}\left(\frac{\tau}{\alpha}+r_{\rm min}\right)^{k-1}e^{\left(-\left[\tau/\alpha+r_{\rm min}\right]/\theta\right)}\exp\left(-\frac{\left[\alpha-\langle\alpha\rangle\right]^{2}}{2\sigma^{2}_{\alpha}}\right)\, ,
\end{eqnarray}
where $p(r,\alpha)=p(r)p(\alpha)$.  Numerically computing the integration in Equation~(\ref{eqn:ptau_re}), we reconstruct $p(\tau)$ first from $p(r_{50})$, 
which are shown in Figure~\ref{fig:ptau_re}. For these plots, the two characteristic parameters of the Gamma distribution, $\{k,\theta\}$, are set at the 
best-fit values (see Table~\ref{tab:fit}) determined in Section~\ref{sec:psize}, while the mean slope and standard deviation of the linear $r$--$\tau$ correlations, 
$\{\langle\alpha\rangle, \sigma_{\alpha}\}$, are set at the numerical values determined by~\citet{ML25} (given in the fifth column of Table~\ref{tab:fit}). 
Comparing the reconstructed $p(\tau)$ with the real distributions that \citet{ML24b} determined at $z=127$ by directly measuring the misalignments between  
protogalactic inertia and initial tidal tensors~\citep[see the top-left panel of Figure 1 in][]{ML24b}, we find that the reconstructed distributions indeed reproduce 
excellently the shapes, behaviors and $scale$-dependences of the real distributions.

It is worth mentioning here that reconstructing $p(\tau)$ in three different $m_{\star}$-ranges is equivalent to reconstructing them on three 
different smoothing scales, since the protogalactic angular momenta of the galaxies belonging to a different $m_{\star}$-range are generated 
by the initial tidal field smoothed on a different scale. As shown in \citet{ML24b}, the shape-variation of $p(\tau)$ is caused by the {\it differences} 
between the smoothing scale of the initial tidal field and the Lagrangian radius of protogalactic halos, but not by their individual values.
A larger difference between the two scales has an effect of broadening $p(\tau)$ and  shifting $\tau_{\rm max}$ to a higher-value where 
$\tau_{\rm max}\equiv \mathop{\rm argmax}_{\tau}p(\tau)$. 

Therefore, the variation of $p(\tau)$ with smoothing scales in a single $m_{\star}$-range is expected to be the same as its variation with $m_{\star}$ on 
a single smoothing scale. This logic explains why the reconstructed distributions, $p(\tau)$, shown in Figure~\ref{fig:ptau_re} match the real distributions 
obtained on three different scales in a single $m_{\star}$-range, although we use the values of $\{\langle\alpha\rangle, \sigma_{\alpha}\}$ determined 
at three different $m_{\star}$ ranges on a single smoothing scale. 
We also reconstruct $p(\tau)$ from $p(r_{90})$ via the same procedure and obtain the identical results in each $m_{\star}$-range. This result implies 
that both of $p(r_{50})$ and $p(r_{90})$ are equally powerful for the reconstruction of the same $p(\tau)$ when the corresponding values of $\{k,\theta\}$, 
$\langle\alpha\rangle$ and $\sigma_{\alpha}$ are appropriately plugged in Equation~(\ref{eqn:ptau_re}). 


\section{Summary and Discussion}\label{sec:con}

Deriving the probability density distributions of the galaxy half-light and $90\%$-light sizes, $\{p(r_{50}),\ p(r_{90})\}$, by analyzing the 
NASA-Sloan Atlas catalog, we have for the first time discovered that the observationally obtained $\{p(r_{50}),\ p(r_{90})\}$ follow 
the bimodal Gamma mixture model, refuting the conventionally accepted log-normal model~\citep[e.g.,][]{she-etal03}. 
It has also been found that the degree of bimodality of $\{p(r_{50}),\ p(r_{90})\}$ depends on the ratio, $r_{50}/r_{90}$,
and that the late-type galaxies with lower concentrations of $r_{50}/r_{90}\ge 0.45$ are best-described by the unimodal Gamma model.  
Whereas, the early-type galaxies with higher concentrations of $r_{50}/r_{90} < 0.45$ have turned out to possess distinct bimodal features, 
which implies that the relaxation process after the latest major merging should be responsible for the observed 
bimodality of galaxy optical size distributions.

In light of the recent numerical result of \citet{ML25}, we have assumed the existence of a linear causal correlation between the optical sizes of late-type 
galaxies and primordial spin factor ($\tau)$ defined as the degree of misalignments between protogalactic inertia and initial tidal tensors. In the linear 
tidal torque theory,  the factors, $\tau$, are directly proportional to the magnitudes of protogalactic angular momenta~\citep{whi84,CT96,LP00}. 
Under this assumption, we have reconstructed the probability density distributions, $p(\tau)$, from the observed $\{p(r_{50}),\ p(r_{90})\}$ of the late-type 
galaxies in three different stellar-mass ranges. It has been revealed that the reconstructed $p(\tau)$ reproduces excellently the overall shapes and scale-dependent 
trends of the real distributions computed at the protogalactic stages in the previous numerical work of \citet{ML24b}. 
This result stands the first observational hint for the existence of a causal correlation between the primordial spin factor and optical sizes of late-type galaxies. 

Although the main focus of the current work has been the reconstruction of $p(\tau)$, our result implies that it should in principle be possible to 
reconstruct the whole $\tau$-fields, $\tau({\bf x})$, on galactic scales from the observable galaxy optical size fields by employing the iterative reconstruction 
algorithm as in \citet{LP00,LP01}. The algorithm was originally developed  for the reconstruction of the initial tidal and density fields from the spin axes. 
The galaxy optical sizes have a couple of advantages over the galaxy spin axes for the reconstruction of initial conditions. 
First, the former is a more readily measurable quantity than the latter, the determination of which is well known to suffer from a degeneracy between the 
clock-wise and counter clock-wise directions~\citep{lee11}. Second, unlike the galaxy spin axes that can reconstruct only the rescaled version of 
initial tidal fields having unity amplitudes~\citep{LP01}, the galaxy optical sizes can reconstruct the full version of initial tidal fields since they are 
connected with the {\it magnitudes} of protogalactic angular momenta. Our future work is in the direction of reconstructing the initial conditions from 
the optical sizes of local late-type galaxies and of investigating if and how the reconstructed initial conditions can constrain the early universe physics.
We hope to report the result elsewhere in the near future.

\acknowledgments

Funding for SDSS-III has been provided by the Alfred P. Sloan Foundation, the Participating Institutions, the National Science Foundation, and the U.S. Department of Energy. The SDSS-III web site is http://www.sdss3.org.
SDSS-III is managed by the Astrophysical Research Consortium for the Participating Institutions of the SDSS-III Collaboration including the University 
of Arizona, the Brazilian Participation Group, Brookhaven National Laboratory, University of Cambridge, University of Florida, the French Participation 
Group, the German Participation Group, the Instituto de Astrofisica de Canarias, the Michigan State/Notre Dame/JINA Participation Group, Johns 
Hopkins University, Lawrence Berkeley National Laboratory, Max Planck Institute for Astrophysics, New Mexico State University, New York University, 
Ohio State University, Pennsylvania State University, University of Portsmouth, Princeton University, the Spanish Participation Group, University of 
Tokyo, University of Utah, Vanderbilt University, University of Virginia, University of Washington, and Yale University.
The GALEX is a NASA Small Explorer. The mission was developed in cooperation with the Centre National d'Etudes 
Spatiales of France and the Korean Ministry of Science and Technology.

We are grateful to an anonymous referee for providing several insightful comments.
JSM acknowledges the support by the National Research Foundation (NRF) of Korea grant funded by the Korean 
government (MEST) (No. 2019R1A6A1A10073437).
JL acknowledges the support by Basic Science Research Program through the NRF of Korea funded 
by the Ministry of Education (RS-2025-00512997).

\clearpage
\clearpage
\clearpage
\clearpage
\clearpage
\begin{deluxetable}{ccccccc}
\tablewidth{0pt}
\tablecaption{Numbers of the spiral galaxies and best-fit parameters of the analytical model for the size distributions in each mass range}
\setlength{\tabcolsep}{3mm}
\tablehead{$M_{\star}$ & $N_{\rm s}$ & $k$ & $10\theta$ & $10^{2}\langle\alpha\rangle$ &  $10^{2}\sigma_{\alpha}$ & $r_{\rm min}$ \\ 
$[h^{-1}M_{\odot}]$ & &  &  &  &  & $[h^{-1}\,{\rm kpc}]$}
\startdata
$[9.5,\ 10]$ & 4824 & $7.98 (8.31)$ & $5.8(11.2)$  &  $2.5 (1.25)$ & $1.0(0.5)$& $1.02 (2.13)$   \\
$[10,\ 10.5]$ & 5836 &  $12.42 (13.55)$ &  $4.7(8.6)$  & $2.5 (1.25)$ & $1.8(1.5)$  &  $1.36 (2.59)$ \\
$[10.5,\ 11]$& 5719 &  $16.36 (17.76)$ & $5.0(9.5)$ & $2.5 (1.25)$ & $1.8(1.8)$ & $2.28 (4.5)$   \\
\enddata
\label{tab:fit}
\end{deluxetable}


\clearpage
\begin{thebibliography}{}
\bibitem[Alam et al.(2015)]{sdssdr12} 
Alam, S., Albareti, F.~D., Allende Prieto, C., et al.\ 2015, \apjs, 219, 1, 12. doi:10.1088/0067-0049/219/1/12
\bibitem[Albareti et al.(2017)]{sdssdr13} 
Albareti, F.~D., Allende Prieto, C., Almeida, A., et al.\ 2017, \apjs, 233, 2, 25. doi:10.3847/1538-4365/aa8992
\bibitem[Bianchi \& GALEX Team(1999)]{galax} 
Bianchi, L. \& GALEX Team\ 1999, \memsai, 70, 365. 
\bibitem[Blanton et al.(2011)]{bla-etal11} 
Blanton, M.~R., Kazin, E., Muna, D., et al.\ 2011, \aj, 142, 1, 31. doi:10.1088/0004-6256/142/1/31
\bibitem[Bullock et al.(2001)]{bul-etal01} 
Bullock, J.~S., Dekel, A., Kolatt, T.~S., et al.\ 2001, \apj, 555, 1, 240. doi:10.1086/321477
\bibitem[Catelan \& Theuns(1996)]{CT96} 
Catelan, P. \& Theuns, T.\ 1996, \mnras, 282, 436. doi:10.1093/mnras/282.2.436
\bibitem[Cadiou et al.(2022)]{cad-etal22} 
Cadiou, C., Pontzen, A., \& Peiris, H.~V.\ 2022, \mnras, 517, 3, 3459. doi:10.1093/mnras/stac2858
\bibitem[Doroshkevich(1970)]{dor70} 
Doroshkevich, A.~G.\ 1970, Astrofizika, 6, 581. 
\bibitem[Finkbeiner et al.(2016)]{fin-etal16} 
Finkbeiner, D.~P., Schlafly, E.~F., Schlegel, D.~J., et al.\ 2016, \apj, 822, 2, 66. doi:10.3847/0004-637X/822/2/66
\bibitem[Forero-Romero et al.(2014)]{for-etal14} 
Forero-Romero, J.~E., Contreras, S., \& Padilla, N.\ 2014, \mnras, 443, 1090
\bibitem[Jiang et al.(2019)]{jia-etal19} 
Jiang, F., Dekel, A., Kneller, O., et al.\ 2019, \mnras, 488, 4, 4801. doi:10.1093/mnras/stz1952
\bibitem[Land et al.(2008)]{zoo1} 
Land, K., Slosar, A., Lintott, C., et al.\ 2008, \mnras, 388, 4, 1686. doi:10.1111/j.1365-2966.2008.13490.x
\bibitem[Lee \& Pen(2000)]{LP00} 
Lee, J. \& Pen, U.-L.\ 2000, \apjl, 532, L5. 
\bibitem[Lee \& Pen(2001)]{LP01} 
Lee, J. \& Pen, U.-L.\ 2001, \apj, 555, 106. doi:10.1086/321472
\bibitem[Lee(2011)]{lee11} Lee, J.\ 2011, \apj, 732, 2, 99. doi:10.1088/0004-637X/732/2/99
\bibitem[Lee et el.(2020)]{lee-etal20} 
Lee, J., Libeskind, N.~I., \& Ryu, S.\ 2020, \apjl, 898, L27
\bibitem[Lee \& Libeskind(2020)]{LL20} 
Lee, J. \& Libeskind, N.~I.\ 2020, \apj, 902, 22. doi:10.3847/1538-4357/abb314
\bibitem[Lee \& Moon(2023)]{LM23} 
Lee, J. \& Moon, J.-S.\ 2023, \apjl, 951, 2, L26. doi:10.3847/2041-8213/acdd75
\bibitem[Lintott et al.(2008)]{zoo2} 
Lintott, C.~J., Schawinski, K., Slosar, A., et al.\ 2008, \mnras, 389, 3, 1179. doi:10.1111/j.1365-2966.2008.13689.x
\bibitem[Masters et al.(2010)]{mas-etal10} 
Masters, K.~L., Nichol, R., Bamford, S., et al.\ 2010, \mnras, 404, 2, 792. doi:10.1111/j.1365-2966.2010.16335.x
\bibitem[Moon \& Lee(2023)]{ML23} 
Moon, J.-S. \& Lee, J.\ 2023, \apj, 952, 2, 101. doi:10.3847/1538-4357/acd9ac
\bibitem[Moon \& Lee(2024a)]{ML24a} 
Moon, J.-S. \& Lee, J.\ 2024a, \jcap, 2024, 5, 111. doi:10.1088/1475-7516/2024/05/111
\bibitem[Moon \& Lee(2024b)]{ML24b} 
Moon, J.-S. \& Lee, J.\ 2024b, \apj, 966, 1, 100. doi:10.3847/1538-4357/ad3825
\bibitem[Moon \& Lee(2025)]{ML25} 
Moon, J.-S. \& Lee, J.\ 2025, \jcap, 2025, 3, 018. doi:10.1088/1475-7516/2025/03/018
\bibitem[Motloch et al.(2021)]{mot-etal21} 
Motloch, P., Yu, H.-R., Pen, U.-L., et al.\ 2021, Nature Astronomy, 5, 283. doi:10.1038/s41550-020-01262-3
\bibitem[Motloch et al.(2022)]{mot-etal22} 
Motloch, P., Pen, U.-L., \& Yu, H.-R.\ 2022, \prd, 105, 8, 083512. doi:10.1103/PhysRevD.105.083512
\bibitem[Oh et al.(2013)]{oh-etal13} 
Oh, K., Choi, H., Kim, H.-G., et al.\ 2013, \aj, 146, 6, 151. doi:10.1088/0004-6256/146/6/151
\bibitem[Park \& Choi(2005)]{PC05} 
Park, C. \& Choi, Y.-Y.\ 2005, \apjl, 635, 1, L29. doi:10.1086/499243
\bibitem[Porciani et al.(2002a)]{por-etal02a} 
Porciani, C., Dekel, A., \& Hoffman, Y.\ 2002a, \mnras, 332, 325. doi:10.1046/j.1365-8711.2002.05305.x
\bibitem[Porciani et al.(2002b)]{por-etal02b} 
Porciani, C., Dekel, A., \& Hoffman, Y.\ 2002b, \mnras, 332, 339. doi:10.1046/j.1365-8711.2002.05306.x
\bibitem[Shen et al.(2003)]{she-etal03} 
Shen, S., Mo, H.~J., White, S.~D.~M., et al.\ 2003, \mnras, 343, 3, 978. doi:10.1046/j.1365-8711.2003.06740.x
\bibitem[Sheng et al.(2023)]{she-etal23} 
Sheng, M.-J., Yu, H.-R., Li, S., et al.\ 2023, \apj, 943, 2, 128. doi:10.3847/1538-4357/acae92
\bibitem[Sheng et al.(2024)]{she-etal24} 
Sheng, M.-J., Zhu, L., Yu, H.-R., et al.\ 2024, \prd, 109, 12, 123548. doi:10.1103/PhysRevD.109.123548
\bibitem[Shim et al.(2024)]{shi-etal24} 
Shim, J., Pen, U.-L., Yu, H.-R., et al.\ 2024, , arXiv:2406.06080. doi:10.48550/arXiv.2406.06080
\bibitem[Teklu et al.(2015)]{tek-etal15} 
Teklu, A.~F., Remus, R.-S., Dolag, K., et al.\ 2015, \apj, 812, 1, 29. doi:10.1088/0004-637X/812/1/29
\bibitem[Vitvitska et al.(2002)]{vit-etal02} 
Vitvitska, M., Klypin, A.~A., Kravtsov, A.~V., et al.\ 2002, \apj, 581, 2, 799. doi:10.1086/344361
\bibitem[Wake et al.(2017)]{wak-etal17} 
Wake, D.~A., Bundy, K., Diamond-Stanic, A.~M., et al.\ 2017, \aj, 154, 3, 86. doi:10.3847/1538-3881/aa7ecc
\bibitem[Wang et al.(2016)]{elucid1} 
Wang, H., Mo, H.~J., Yang, X., et al.\ 2016, \apj, 831, 2, 164. doi:10.3847/0004-637X/831/2/164
\bibitem[Wang et al.(2014)]{elucid2} 
Wang, H., Mo, H.~J., Yang, X., et al.\ 2014, \apj, 794, 1, 94. doi:10.1088/0004-637X/794/1/94
\bibitem[Wechsler et al.(2002)]{wec-etal02} 
Wechsler, R.~H., Bullock, J.~S., Primack, J.~R., et al.\ 2002, \apj, 568, 1, 52. doi:10.1086/338765
\bibitem[White(1984)]{whi84} 
White, S.~D.~M.\ 1984, \apj, 286, 38. doi:10.1086/162573
\bibitem[Yu et al.(2019)]{yu-etal19} 
Yu, H.-R., Pen, U.-L., \& Wang, X.\ 2019, \prd, 99, 12, 123532. doi:10.1103/PhysRevD.99.123532
\bibitem[Yu et al.(2020)]{yu-etal20} 
Yu, H.-R., Motloch, P., Pen, U.-L., et al.\ 2020, \prl, 124, 10, 101302. doi:10.1103/PhysRevLett.124.101302
\end{thebibliography}
\end{document}